\documentclass[10pt,letterpaper]{article}
\usepackage{opex3}

\begin{document}


\title{Waveguide-based single-shot temporal cross-correlator}


\author{Moti Fridman$^1$, Yoshitomo Okawachi$^2$, St\'ephane Clemmen$^2$, Micha\"{e}l M\'{e}nard$^2$, Michal Lipson$^2$, and Alexander L. Gaeta$^2$}

\address{
$^1$ Faculty of Engineering, Bar Ilan University, Ramat Gan 52900, Israel \\
$^2$ School of Applied and Engineering Physics, Cornell University, Ithaca, NY 14853, USA
}

\begin{abstract}
We describe a novel technique for performing a single-shot optical cross-correlation in nanowaveguides. Our scheme is based on four-wave mixing between two orthogonally polarized input signals propagating with different velocities due to polarization mode dispersion. The cross-correlation is determined by measuring the spectrum of the idler wave generated by the four-wave mixing process.
\end{abstract}

\ocis{070.4550 (Correlators), 070.4340 (Nonlinear optical signal processing), 230.7370 (Waveguides)}


\section{Introduction}

Cross-correlation gives a measure for the similarity between two input signals, therefore, it is implemented in numerous data processing systems~\cite{optica_routers, optic_router2, jalali_camera, jalali_corelator, frog, mechanic, freespace, singleshot, Xray}. Today, the calculation of the cross-correlation is done in an electronic processing unit~\cite{electronics} and since most of the data communication is done via optical fibers, an optics-to-electronics convertor is needed. This convertor is slow compared to the flow of information in the optical fibers and imposes an electronic bottleneck on the optical communication. Optical routers can eliminate the bottleneck and a key element in optical routers is the optical cross-correlation device for header detection~\cite{optica_routers, optic_router2}. Optical cross-correlation is also needed for pattern recognition in triggering ultrafast cameras~\cite{jalali_camera, jalali_corelator}. Both optical routers and ultrafast camera triggers require cross-correlation of non-repetitive signals and in both cases the cross-correlator should be embedded in a fiber or a waveguide. Optical cross-correlation is also a commonly used technique for ultrashort pulse characterization~\cite{frog} and developing a technique which does not require repetitive pulses and can rather analyze a single-short pulse will make high impact on the ultrafast lasers community. Many approaches for optical cross-correlation have been developed, including mechanically moving filters~\cite{mechanic}, diffraction in free space~\cite{freespace, Xray} and nonlinear based approaches~\cite{singleshot, Xray}. However, all these techniques are slow, performed in free space or require repetitive signals.

Here we present, an ultrafast optical cross-correlation technique that utilizes the nonlinearity and polarization mode dispersion (PMD) of a nanowaveguide to perform a single-shot cross-correlation. We note that the PMD rises from the birefringence of the waveguide which is engineered by manipulating the cross-section of the waveguide. We show that measurement of the idler wave spectrum generated from the interaction of two orthogonally-polarized chirped input waves (signal and pump) via four-wave mixing (FWM) yields the cross-correlation of the two inputs.

\section{Theoretical background}
Our technique for performing the cross-correlation is based on the FWM process within a waveguide between the pump and the signal waves with amplitudes $A_p$ and $A_s$, respectively. This interaction yields an idler wave with an amplitude $A_i(t)$, given by
\begin{equation}
A_i(t) \propto A_s(t)^{*} A_p^2(t).
\end{equation}
Integrating over time to obtain the measured intensity $I_i$ by slow detectors, yields
\begin{equation}
I_i = \int A_i^2(t) dt \propto \int \left( A_s(t)^{*} A_p^2(t) \right)^2 dt.
\end{equation}
Thus, the measured amplitude of the idler wave, which is equal to square root of $I_i$, is proportional to the correlation between the signal wave and the square of the pump wave. To find the correlation between an input signal and a temporal filter, the pump wave needs to be tailored as square root of the desired filter. The resulting amplitude of the idler wave equals to the correlation between the input signal and the filter.

A measurement of cross-correlation requires the input signal to pass through the other input signal while measuring the correlation between them at each time. This is accomplished by rotating the polarization of the pump wave with respect to the signal wave such that the polarization of the two beams is orthogonal to each other. In the presence of PMD in the waveguide, the signal wave and the pump wave travel with different group velocities due to birefringence so one passes through the other.

We inject the signal ahead of the pump wave with an initial temporal separation $\Delta t$. As the two waves propagate through the waveguide, the pump passes through the signal wave so $\Delta T$ changes. Due to FWM interaction between the pump and the signal waves an idler wave is generated at each $\Delta t$. The idler amplitude as a function of time is proportional to the correlation as a function of $\Delta t$. To obtain the cross correlation, we need to measure the idler intensity as a function of time~\cite{FWM1, FWM2}, however, this is not possible for an ultrashort signal wave of with fluctuations of less than a picosecond. To overcome this problem, we map time-to-frequency by suitably chirping the signal and pump waves in such a way that the frequency of the idler wave is a function of the temporal delay between the signal and the pump waves. Thus, for different temporal separations the generated idler wave has different frequency.

We illustrate this concept by assuming that the frequency of the chirped signal wave as a function of time expressed as,
\begin{equation}
\omega_s(t)=\omega_{s0} + \alpha (t+\Delta t),
\end{equation}
where $\omega_{s0}$ is the central frequency of the signal wave, and $\alpha$ is the slope of the chirping. Since the frequency of the resulting idler is
\begin{equation}\label{EqOmega}
\omega_i = 2 \omega_p - \omega_s,
\end{equation}
the appropriate slope of the chirped pump should be $\alpha/2$ such that,
\begin{equation}
\omega_p(t)= \omega_{p0} + \frac{\alpha}{2} t,
\end{equation}
where $\omega_{p0}$ is the central frequency of the pump wave. The frequency of the generated idler is then,
\begin{equation}\label{EqDelta}
\omega_i(t)=2 \omega_{p0} - \omega_{s0} - \alpha \Delta t.
\end{equation}
It can be seen from Eqs. (2)-(4), that although the signal and the pump waves are broadband and chirped (their frequencies depend linearly on $t$), the bandwidth of the generated idler wave is narrowband. Also, the frequency of the idler wave is constant in time and depends only on the temporal separation between the signal and the pump waves. For a temporal separation $\Delta t$ between the signal and the pump waves, Eq. (1) yields
\begin{equation}\label{EqTotal}
A_i(\Delta t) \propto \int A_s(\tau + \Delta t) A_p^2(\tau) d \tau.
\end{equation}
Combining Eq.~\ref{EqDelta} into the left hand side of Eq.~\ref{EqTotal} together with Eq.~\ref{EqOmega} gives,
\begin{equation}
A_i\left(\frac{\omega_i-\omega_{i0}}{\alpha}\right) \propto \int A_s(\tau + \Delta t) A_p^2(\tau) d \tau,
\end{equation}
where $\omega_{i0}$ is the central frequency of the idler wave and $A_i(\omega_i)$ is the spectrum of the idler wave. Thus, the cross-correlation of the signal wave with the pump wave is proportional to the scaled spectrum of the idler wave. Therefore, by correctly chirping the pump and the signal waves and utilizing the polarization mode dispersion, the spectrum of the resulting idler wave is proportional to the cross-correlation between the signal wave and the square of the pump wave.

\section{Experimental results of the optical cross-correlator}
\begin{figure}[htb]
\centerline{\includegraphics[width=6cm]{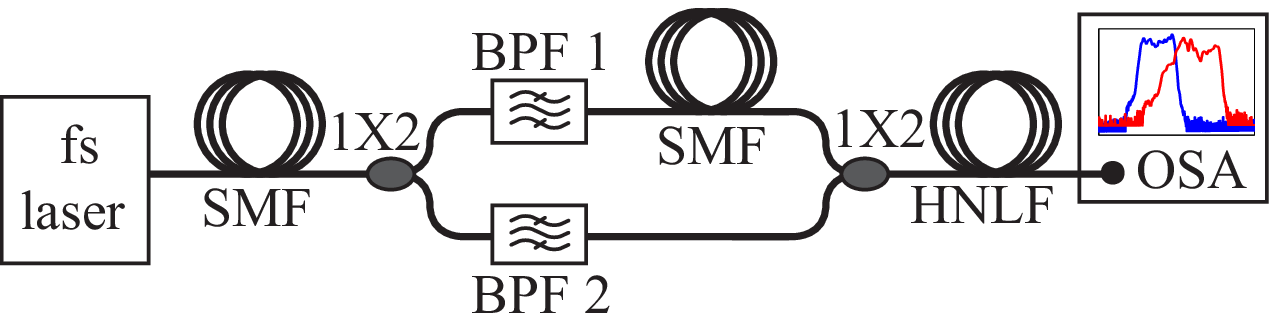} \includegraphics[width=3cm]{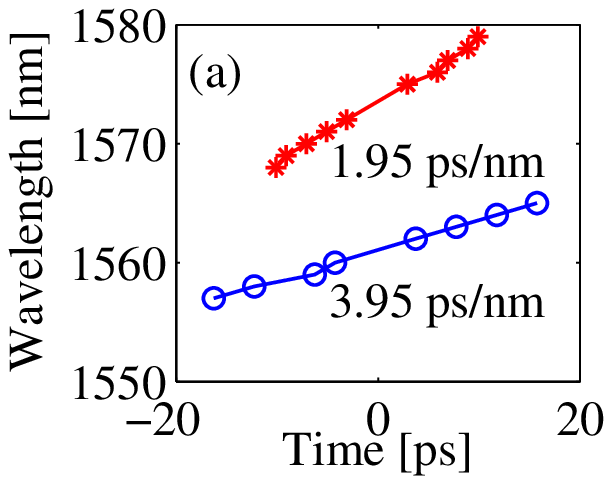}}
\caption{\label{systemHNLF}Left: Schematic of the experimental configuration for measuring the idler generated by the chirped signal and pump. HNLF - highly nonlinear fiber; 1X2 - 50/50 beam splitter; OSA - optical spectrum analyzer; SMF - 110 m of single mode fiber; BPF - band-pass filter, where the center wavelength of BPF 1 is 1561$\pm$4 nm, and BPF 2 is 1573.5$\pm$6 nm~\cite{explain}. Right: Measurements of the wavelength as a function of time for the pump (circles) and the signal (asterisks).}
\end{figure}

To verify the basic concept of our approach, we demonstrate that the generated idler from appropriately chirped broadband signal and pump waves is narrow-band and its frequency is a function of the time delay between the pump and the signal. The experimental configuration for measuring the spectrum of the idler generated from the chirped pump and signal is presented in Fig.~\ref{systemHNLF}. An ultra-short pulse is stretched in time with a 110-m length of single-mode fiber (SMF) such that it is chirped by 1.95 ps/nm. The pulse is then split into two separate arms that contain bandpass filters to produce a signal at $1573.5 \pm 6$ nm and a pump at $1561 \pm 4$ nm with peak power of about $100 mW$ for both waves. The pump is stretched in time using another 110 m of SMF to 3.95 ps/nm. Using a tunable narrow band filter followed by a fast (30 GHz) photo-detector connected to a sampling scope, we measured the exact arrival time for each wavelength to obtain a spectrogram of both the pump and the signal waves [Fig.~\ref{systemHNLF}(a)].

\begin{figure}[htbp]
\centerline{\includegraphics[width=7.5cm]{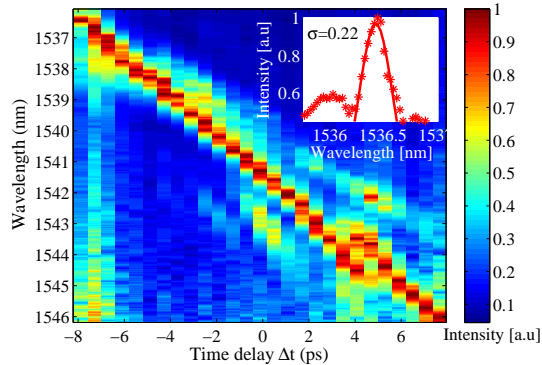}}
\caption{\label{Spectrum2mHNLF}Spectra of the idler for different time delays between the pump and the signal. Inset - typical spectrum at a specific time delay with a Gaussian fit. The width of the gaussian fit is 0.22 nm.}
\end{figure}

We inject both waves into a short section of highly nonlinear fiber (HNLF) (CorActive SCF-UN-3/125-25) and measure the spectrum of the resulting idler wave as the time delay between the pump and the signal waves is varied. The spectra are shown in Fig.~\ref{Spectrum2mHNLF}, where red denotes high intensity and blue denotes low intensity. The inset shows a single spectrum at a specific time delay, and its calculated width is 0.22 nm. Here we use 2 m of HNLF, which in terms of nonlinearity is equivalent to a 0.3-mm-long silicon waveguide, the idler is clearly detected for all time delays (from -8.5 ps to 8.5 ps), and the width of the peak is less than 0.25 nm. We attribute the noise at the long wavelength region to be due to self phase modulations of the pump wave. This noise can be reduced by using signal and pump waves that have a greater separation in the spectral domain.

\begin{figure}[htb]
\centerline {\includegraphics[width=7.9cm]{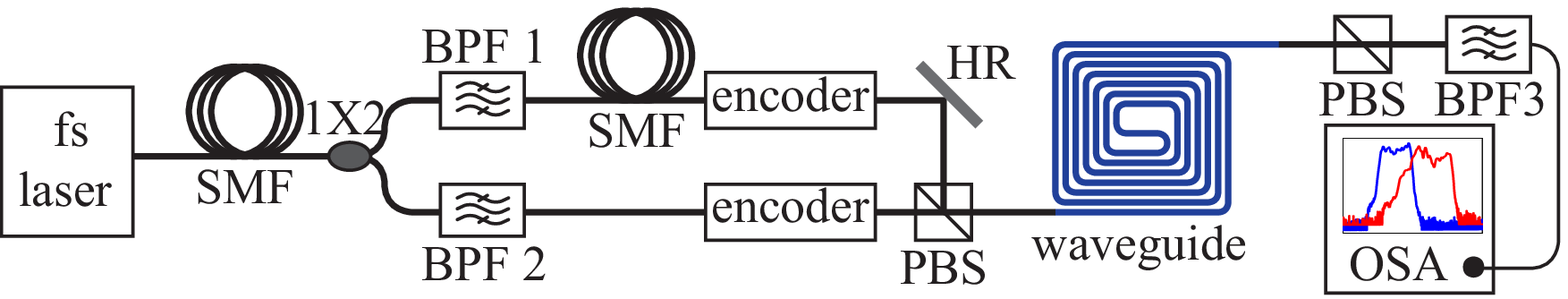} }
\centerline {
            \includegraphics[width=3.5cm]{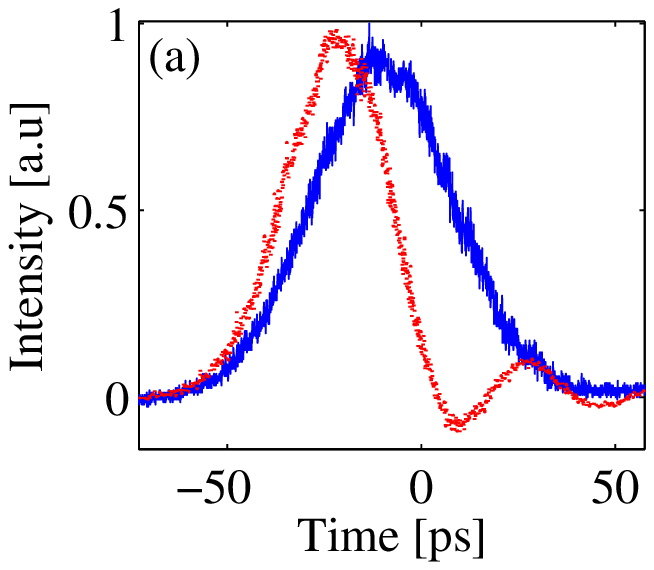}
            \includegraphics[width=3.5cm]{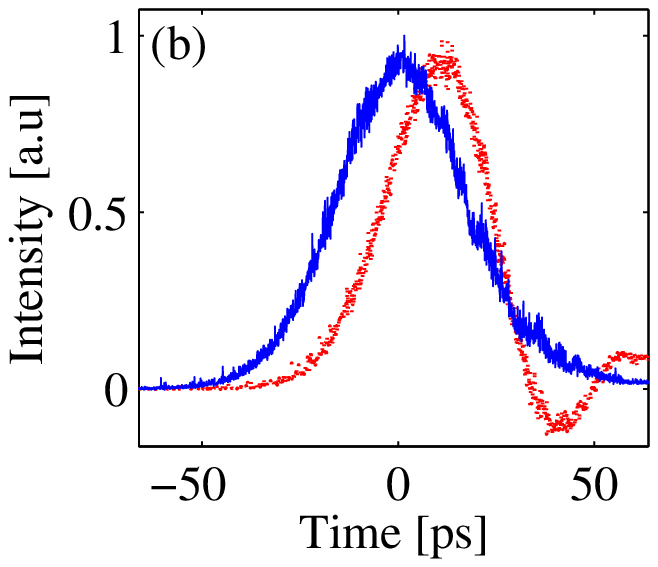}
            }
\caption{\label{system}Top: Schematic of the experimental configuration for the optical cross-correlator in the silicon nanowaveguide. PBS - polarization beam splitter; 1X2 - 50/50 beam splitter; HR - high reflection mirror; OSA - optical spectrum analyzer; SMF - 110 m of single mode fiber; BPF - band-pass filter, where the center wavelength of BPS 1 is 1561$\pm$4 nm, BPS 2 is 1573.5$\pm$6 nm, and PMS 3 is 1545$\pm$10 nm~\cite{explain}. (a) temporal measurements of the pump wave (solid curve) and the signal wave (dashed curve) before the waveguide; (b) temporal measurements of the pump wave (solid curve) and the signal wave (dashed curve) after the waveguide.}
\end{figure}

\begin{figure}[htb]
\centerline{\includegraphics[width=7cm]{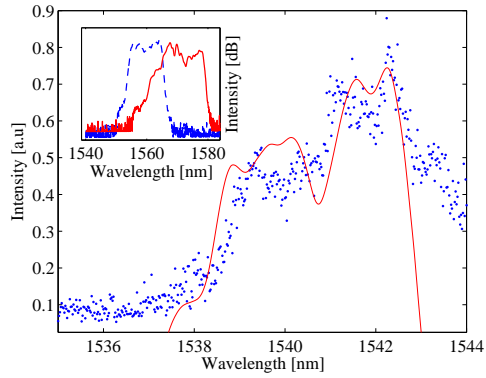}}
\caption{\label{crosscorr}Experimental results of the idler spectrum (dots) together with the calculated cross-correlation of the pump with the signal. Inset shows the spectrum of the pump wave (dashed curve) and the spectrum of the signal (solid curve).}
\end{figure}

Next, we present that the cross-correlation between two arbitrary input functions can be determined by measuring the spectrum of the generated idler wave. The input functions are encoded on both the signal and pump waves using fiber Bragg gratings that are serially connected. In order to account for the square of the pump in Eq. (1), we encode on the pump the square root of the input data. In principle, one could use a diffraction grating followed by a spatial light modulator or a nonlinear process to encode the input function in real-time. The experimental configuration is presented in Fig.~\ref{system}. Here the signal and pump waves are orthogonally polarized, and the HNLF is replaced with a silicon nanowaveguide which is birefringent. The waveguide is 1-cm long and has a rectangular cross-section of 710 X 310 nm with a PMD of 18 ps/cm. After encoding the input functions, the pump and the signal waves are combined by a polarization beam combiner and injected into the silicon nanowaveguide. The signal is injected 9 ps before the pump, as shown in Fig.~\ref{system}(a). Due to the PMD of the silicon nanowaveguide, the signal wave propagates slower than the pump wave such that it passes through the pump within the waveguide and exits 9 ps after the pump wave [Fig.~\ref{system}(b)], all these results were acquired in a single shot. We filter the generated idler from the output channel of the nanowaveguide by using a band-pass filter followed by a polarization beam splitter and measure the idler spectrum, as shown in Fig.~\ref{crosscorr} (dotted curve) together with the calculated cross-correlation between the signal and pump waves (solid curve). We attribute the lack of perfect agreement to polarization dependent losses in the waveguide. In our case, the losses where 10 dB/cm which limit the length of the waveguide to no more than few centimeters, also the input and output losses to the silicon waveguide were 5 dB. By designing a waveguide with lower losses it is possible to increase the accuracy of the optical cross-correlator. Therefore, switching to highly nonlinear polarization maintaining fibers, or tightly coiled Chalcogenide glass fibers can obtain the same nonlinearity as silicon waveguide with lower losses and higher PMD.

\section{Experimental demonstration of pattern recognition with optical cross-correlator}
\begin{figure}[htb]
\centerline{\includegraphics[width=3.5cm]{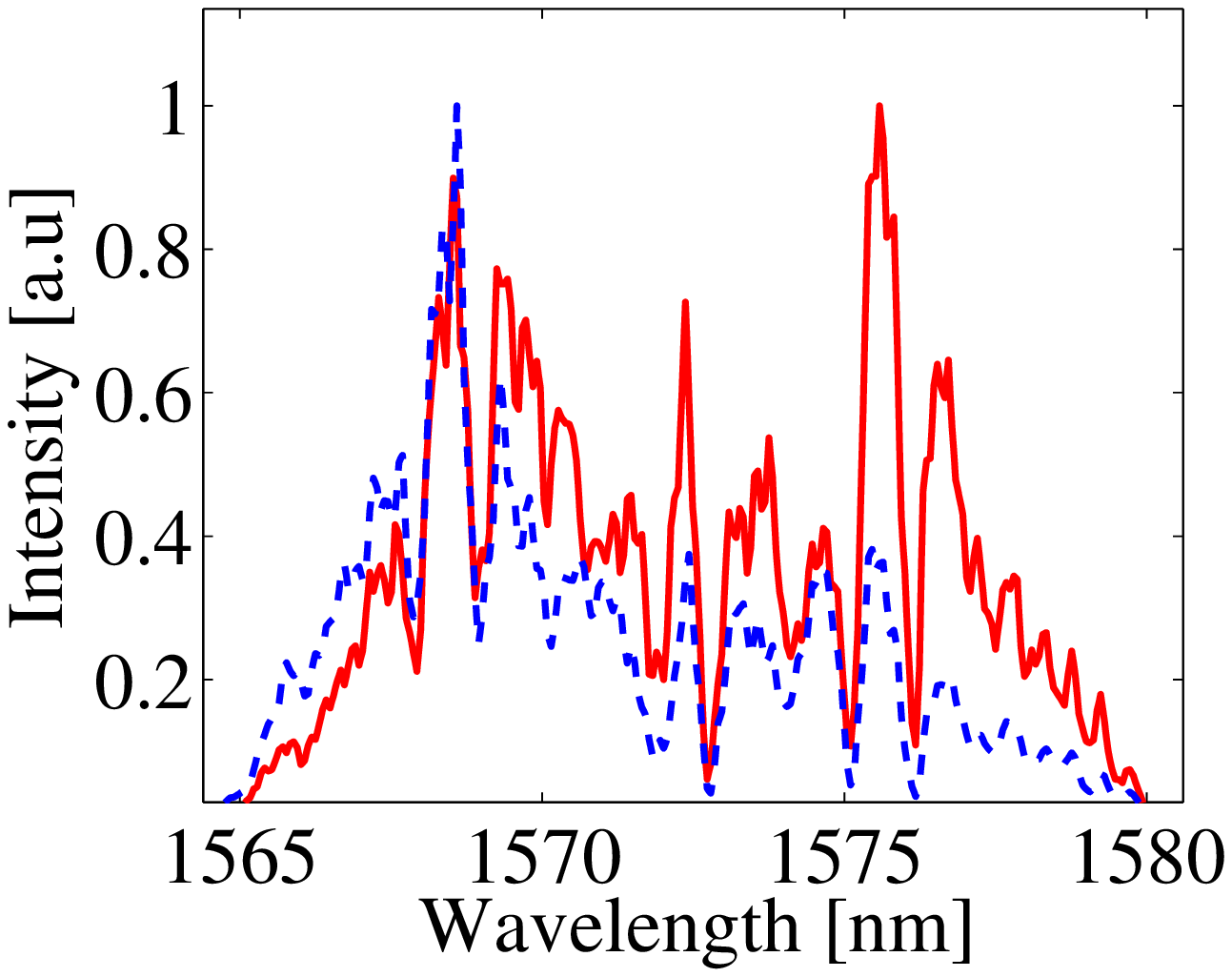}\includegraphics[width=3.5cm]{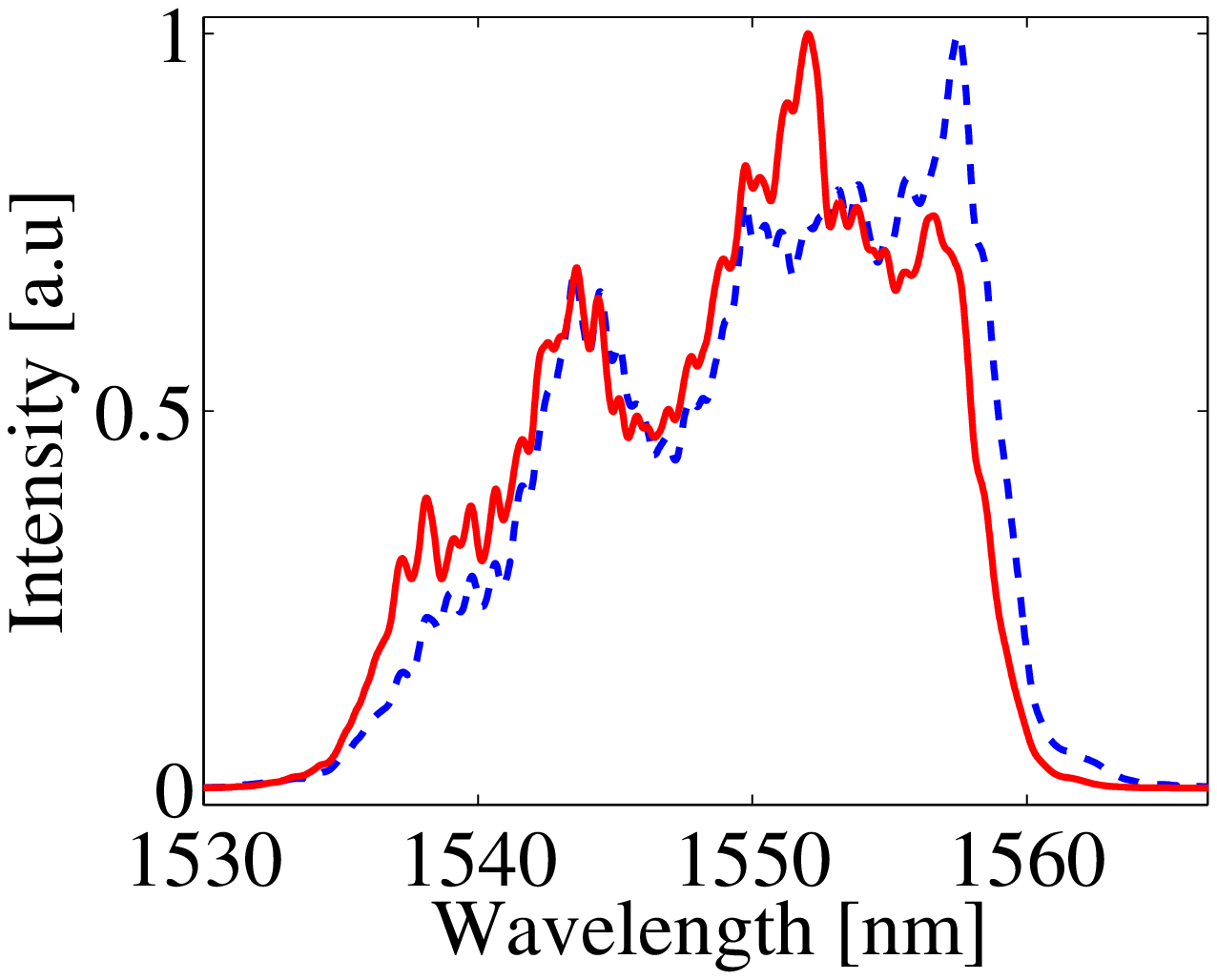}}
\caption{\label{SpectrumForTwoSignal} Left: measured spectra of the two signals. Right: calculated idler spectrum by cross-correlation of the two signals with the pump. Dashed curve: signal $a$, solid curve: signal $b$}
\end{figure}

Finally, we demonstrate a proof-of-principle concept were this system is able to distinguish between two distinct input signals. Each signal is produced using serially-connected fiber Bragg gratings that reflects different wavelength channels from a pulse with a Gaussian shaped spectrum. The resulting spectra are presented in Fig.~\ref{SpectrumForTwoSignal}(left) where the dashed curve denotes signal $a$, and the solid curve denotes signal $b$. We code "0110111010" into signal $a$ and "1111010111" into signal $b$. The temporal length of both signals is less than 4 ps which is equivalent to transfer bit-rate higher than 1 THz, therefore, no electronic detector can distinguished between these two signals. We tailor the spectrum of a pump wave such that cross-correlation of signals $a$ and $b$ with the same pump wave results in different idler waves, so the two signals can be distinguished using our optical cross-correlator. The calculated cross-correlation of the pump wave with each signal waves is presented in Fig.~\ref{SpectrumForTwoSignal}(right) where the dashed curve denotes the cross-correlation with signal $a$ and the solid curve with signal $b$. The peak in the calculated cross-correlation of $a$ is displaced compared to the cross-correlation of $b$ by 5.5 nm.

\begin{figure}[htb]
\centerline{\includegraphics[width=7cm]{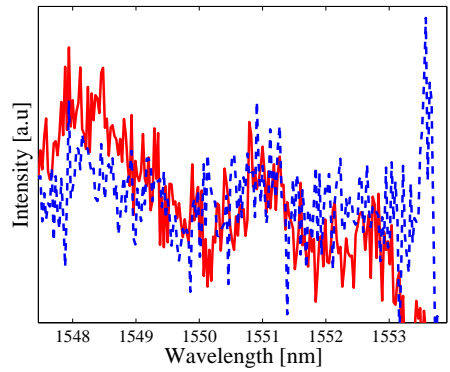}}
\caption{\label{idler4}Measured spectra of the two idler waves created through four wave mixing of the pump with signal waves $a$ (dashed curve) and $b$ (solid curve).}
\end{figure}

We acquire the spectra of the two idler waves resulting from the FWM interaction between the pump wave and signals $a$ and $b$, and the results are presented in Fig.~\ref{idler4}. The peaks in the spectra are displaced by 5.7 nm which is close to the calculated result of 5.5 nm. We attribute the discrepancy between the calculated idler spectrum shift and the measured idler spectrum to inaccuracies in realigning the system after switching from signal $a$ to signal $b$ since they are produced in different optical arms. Nevertheless, our system readily distinguishes between these two signals, and the shapes of the idler spectra resemble that of the calculated cross-correlations for both signal waves.

\newpage

\section{Conclusions}
In summary, we demonstrated a system for measuring the single-shot optical cross-correlation between two signals using a nanowaveguide by simply measuring the spectrum of the idler wave generated from FWM process between the pump and the signal waves. By designing a suitable waveguide with high PMD concordantly with low losses the accuracy by which the cross-correlation is measured can be improved.


\begin{thebibliography}
\bibitem{explain} These are only schematics of the experimental configuration symplified for the reader. The experimental setup was fully integrated inside waveguides.
\end{thebibliography}
\end{document}